\documentclass[letter]{aa} 
\usepackage{graphicx}
\usepackage{txfonts}
\usepackage{longtable}
\usepackage{lscape}
\usepackage{natbib}
\bibpunct{(}{)}{;}{a}{}{,} 
\newcommand\kms{{\rm\,km\,s^{-1}}}

\newcommand\teff{T_{\rm eff}}

\begin{document}

\title{
Chemical evolution of the Galactic bulge as traced by\\
microlensed dwarf and subgiant stars\thanks{
Based on observations carried out at the
European Southern Observatory telescopes on Paranal, Chile, 
Program ID 085.B-0399.
}
}

\subtitle{
III. Detection of lithium in the metal-poor bulge dwarf MOA-2010-BLG-285S
}

\titlerunning{Chemical evolution of the Galactic bulge as traced by microlensed dwarf and subgiant stars. III. Li}

\author{
T. Bensby\inst{1,2}
\and
M. Asplund\inst{3}
\and
J.A. Johnson\inst{4}
\and
S. Feltzing\inst{2}
\and
J. Mel\'endez\inst{5}
\and
S. Dong\inst{6}
\and\\
A. Gould\inst{4}
\and
C. Han\inst{7}
\and
D. Ad\'en\inst{2}
\and
S. Lucatello\inst{8}
\and
A. Gal-Yam\inst{9}
 }

\institute{European Southern Observatory, Alonso de Cordova 3107,
Vitacura, Casilla 19001, Santiago 19, Chile
\and
Lund Observatory, Box 43, SE-221\,00 Lund, Sweden
\and
Max Planck Institute for Astrophysics, Postfach 1317, 85741 Garching, Germany
\and
Department of Astronomy, Ohio State University, 140 W. 18th Avenue,
Columbus, OH 43210, USA
\and
Departamento de Astronomia do IAG/USP, Universidade de S\~ao Paulo, 
Rua do Mat\~ao 1226, S\~ao Paulo, 05508-900, SP, Brasil
\and
Institute for Advanced Study, Einstein Drive, Princeton, NJ 08540,
USA
\and
Department of Physics, Chungbuk National University, Cheongju
361-763, Republic of Korea
\and
INAF-Astronomical Observatory of Padova, Vicolo dell'Osservatorio 5,
35122 Padova, Italy
\and
Benoziyo Center for Astrophysics, Weizmann Institute of Science,
76100 Rehovot, Israel
}


\date{Received 13 September 2010 / Accepted xx xxxxx 2010}

\offprints{T. Bensby, \email{tbensby@astro.lu.se}}

  \abstract
  %
   {
   In order to study the evolution of Li in the Galaxy it
   is necessary to observe dwarf or subgiant stars. These are 
   the only long-lived stars whose
   present-day atmospheric chemical composition reflect their
   natal Li abundances according to standard models of stellar evolution.  
   Although Li has been extensively
   studied in the Galactic disk and halo,  to date there is only 
   one uncertain detection of Li in an unevolved bulge star.
   } 
   {
   Our aim with this study
   is to provide the first clear detection of Li in the Galactic 
   bulge, based on an analysis of a dwarf star that has largely retained
   its initial Li abundance.
   }
   {
   We have performed a detailed elemental abundance analysis of 
   the bulge dwarf star MOA-2010-BLG-285S using a high-resolution,
   and high signal-to-noise spectrum obtained with the UVES
   spectrograph at the VLT when the object was optically
   magnified during a gravitational microlensing event 
   (visual magnification $A\sim550$ during observation).
   The lithium abundance was determined through synthetic line 
   profile fitting of the $^{7}$Li resonance doublet line
   at 670.8\,nm. The results have been corrected for departures from LTE.
   }
   {
   MOA-2010-BLG-285S is, at $\rm [Fe/H]=-1.23$, the most metal-poor
   dwarf star detected so far in the Galactic bulge. 
   Its old age (12.5 Gyr) and enhanced $\rm [\alpha/Fe]$ ratios
   agree well with stars in the thick disk at similar metallicity.  
   This star represents the first unambiguous detection of Li in a 
   metal-poor dwarf star in the Galactic bulge. We find an NLTE corrected 
   Li abundance of $\rm\log\epsilon(Li)=2.16$,
   which is consistent with values derived for Galactic disk and halo dwarf
   stars at
   similar metallicities and temperatures.
   }
   {
   Our results show that there are no signs of Li enrichment or production in 
   the Galactic bulge during its earliest phases. Observations of Li
   in other galaxies ($\omega$ Cen) and other components of the Galaxy suggest further that
   the Spite plateau is universal.
   }

   \maketitle

\section{Introduction}\label{sec:intro}

Recent studies have shown agreement between the chemical evolution 
of the Galactic bulge and the thick disk 
\citep{melendez2008,alvesbrito2010,bensby2010letter,bensby2010}.
While many observations of Li in the Galactic halo and the disk(s)
have been carried out, no measurement of the Li abundance
in unevolved bulge stars
have been secured. It is therefore unknown if the bulge Li abundance 
agrees with the Li plateau of the Galactic halo \citep{spite1982},
or if it shows the same types of depletion or enrichment that many 
of the disk stars do. 

Li is the only element besides H and He that is 
produced in measurable amounts in the 
Big Bang with a predicted primordial $^{7}$Li abundance of
$\rm\log\epsilon(Li)\approx2.72\pm0.05$  \citep[e.g.,][]{cyburt2008}.
At later stages, possible production sites for 
Li include cosmic rays \citep[e.g.,][]{reeves1970},
and AGB stars in certain mass ranges \citep[e.g.,][]{sackmann1999}.
However, Li is a fragile element which  is destroyed in the interior of
stars when the temperature exceeds about $2.5 \cdot 10^6$\,K 
\citep[e.g.,][]{weymann1965}. 
The only stars that retain their initial Li surface abundances are 
dwarf and subgiant stars with effective temperatures in the range 
6000 to 6500\,K. 
Cooler dwarf stars have convective zones that mix the 
surface Li down to hotter regions where it is destroyed, while for
hotter dwarf stars, at solar metallicities, Li is again
destroyed \citep[the so-called Li-dip, see, e.g.,][]{boesgaard1986}.
The solar Li abundance is a factor of 160 lower than measured in
the most pristine meteorites \citep{asplund2009},
which reflects a secular Li depletion over the past 4.5\,Gyr
\citep{baumann2010}.
As stars evolve to become red giant branch (RGB) stars, the Li abundance 
decreases due to the first dredge-up and later through extra mixing 
at the RGB bump
\citep[e.g.,][and \citealt{pinsonneault1997} for a general review 
on mixing in stars]{iben1965,charbonnel2007,lind2009b}.
This is also the reason why RGB stars 
cannot be used for Galactic chemical evolution studies of Li 
(although they can be safely used for many other elements).

Being confined to dwarf stars has given an incomplete and
poorly understood picture on the Galactic evolution of Li.
It is only for Galactic halo and disk stars in the solar neighbourhood 
for which
Li has been unambiguously observed in dwarf stars. For instance, 
studies of metal-poor dwarf stars in the Galactic halo have revealed a cosmological Li problem: 
the measured Li abundance is a factor of $2-5$ lower than predicted 
from standard Big Bang nucleosynthesis (BBNS)
\citep[e.g.,][]{spite1982,asplund2006,bonifacio2007,aoki2009,sbordone2010,melendez2010li}.
The reasons for this are still unknown but both non-standard stellar Li depletion 
\citep[e.g.,][]{korn2006} and non-standard BBNS \citep[e.g.,][]{jedamzik2009}
have been invoked.
Studies of the Galactic disk show a wide scatter of Li abundances,
with many stars indicating that Li increases to very high levels
as well as decreasing to very low levels \citep[e.g.,][]{lambert2004}.

Dwarf stars in the bulge are usually too faint in order to get
the high-resolution spectra needed to analyse Li. However, 
in a pioneering study on using microlensing events to study 
bulge dwarf stars, \cite{minniti1998} claimed a detection of Li in 
MACHO-1997-BLG-45/47. Later, \cite{cavallo2003} re-analysed the 
spectrum obtained by \cite{minniti1998} and could not 
confirm the Li detection due to the limited S/N. Furthermore, in \cite{bensby2010} 
who analysed 15 microlensed bulge dwarfs, attempts were made to 
include MACHO-1997-BLG-45/47 as well, but it had to be excluded 
as the spectrum obtained by \cite{minniti1998} was deemed of
insufficient quality for a trustworthy abundance analysis. Hence, 
the Li detection in the bulge by \cite{minniti1998} is uncertain.
A few Li-rich RGB stars in the bulge have been discovered 
\citep{gonzalez2009}. However, as explained above, such stars have 
not retained their original Li abundances
but have experienced internal Li destruction as well as production.

In this Letter we present the result from a microlensing event
toward the Galactic bulge for which we have obtained a high-resolution
spectrum of high quality of the source star. The star is a dwarf star 
in the bulge that is sufficiently hot to not have developed a deep 
convective zone in its atmosphere, meaning that the initial Li 
abundance of the star is intact. 

\section{Observations and lens effects}

MOA-2010-BLG-285S was identified as a possible high-magnification
microlensing event toward the bulge at $(l,\,b)=(0.3,\,-2.6)$\,deg
in early June 2010 with the MOA alert 
system\footnote{\tt https://it019909.massey.ac.nz/moa/alert/index.html} 
\citep[e.g.,][]{bond2001}. The intrinsic source flux (inferred 
from the microlensing model) indicated that the source star was a
dwarf star and we triggered our ToO observations
with the ESO Very Large Telescope (VLT) on Paranal on June 6.  Using 
the UVES spectrograph \citep{dekker2000short}, 
configured with dichroic number 2, the target was observed for a 
total of two hours, split into three 40 minute exposures. The resulting 
spectrum was recorded on three CCDs with wavelength coverages between 
376-498\,nm (blue CCD), 568-750\,nm (lower red CCD), and 
766-946\,nm (upper red CCD). A slitwidth of $1''$  yielded a 
spectral resolution  of $R\approx45\,000$. The data were reduced with 
the UVES pipeline (version 4.4.5). The signal-to-noise 
($S/N$) ratio per pixel at 670\,nm  is $\sim170$.
Before the observation of the main target, a rapidly rotating 
B2V star (HR\,6141) was observed at an airmass similar to what was 
expected for the Bulge star. The featureless spectrum of HR\,6141
was used to divide out telluric lines in the bulge star spectrum. 

\begin{figure}
\resizebox{\hsize}{!}{
\includegraphics[bb=18 175 592 550,clip]{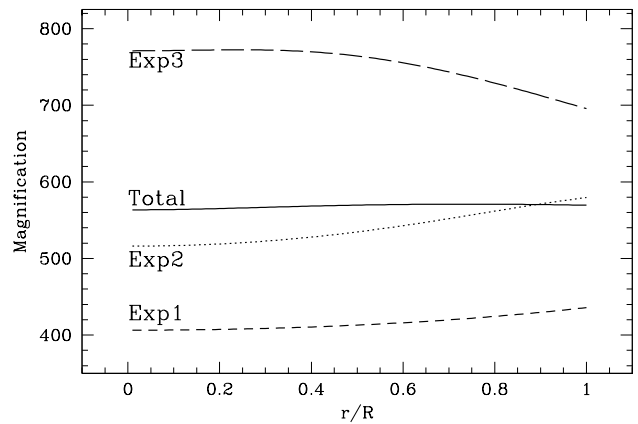}}
\caption{The magnification profiles during the three UVES
exposures of MOA-2010-BLG-285S (dashed and dotted lines), and 
the total magnification 
profile when co-adding all three exposures (solid line).
                  \label{fig:magprofiles}}
\end{figure}

The lens for this event turned out to be in a binary system, and
during the observations the source approached very closely to a cusp
of the binary-lens caustics. This means that
the source star can not be treated as infinitesimally small,
and that there may be substantial differential magnification of the
source's surface, which in turn could conceivably have an impact on the
interpretation of the spectrum. The effect of limb-darkening 
was investigated by \cite{johnson2010} who found that 
the impact of finite source effects on the spectral analysis
is typically very small, unless the spectrum is taken when a
strong cusp or the caustic lies directly over the source.
We have checked the 
effects on the spectra that we obtained of MOA-2010-BLG-285S. As can 
be seen in Fig.~\ref{fig:magprofiles} the magnifications as well
as the magnification profiles changed between the three UVES exposures. 
When adding all three exposures the total magnification is around
550 and almost constant all over the surface of the source.
Hence, the effects on the co-added spectrum are negligible.

\section{Stellar parameters and line synthesis}

Stellar parameters and elemental abundances  were determined in the same way 
as for our previous sample of microlensed dwarf stars in the bulge
\citep[see][]{bensby2009,bensby2010}. Uncertainties in the stellar 
parameters and abundance ratios have been calculated according to the 
prescription in \cite{epstein2010}. 

The Li abundance was determined through line profile
fitting of the \ion{$^7$Li}{i} resonance doublet line at 670.8\,nm. 
This line is highly structured and we have adopted the
$\log gf$-values for the different line components from \cite{smith1998}.
The calculation of synthetic spectra is done with the 
Uppsala {\sc spectrum} software, and our methodology is fully described 
in \cite{bensby2006} where the forbidden [\ion{C}{i}] at 872.7\,nm
was analysed. The difference is that we here analyse a different 
wavelength region and are therefore using a different line to determine 
the width of the RAD-TAN profile; here we use the Fe\,{\sc i} line at 
667.8\,nm.  

\begin{figure}
\resizebox{\hsize}{!}{
\includegraphics[bb=18 150 592 600,clip]{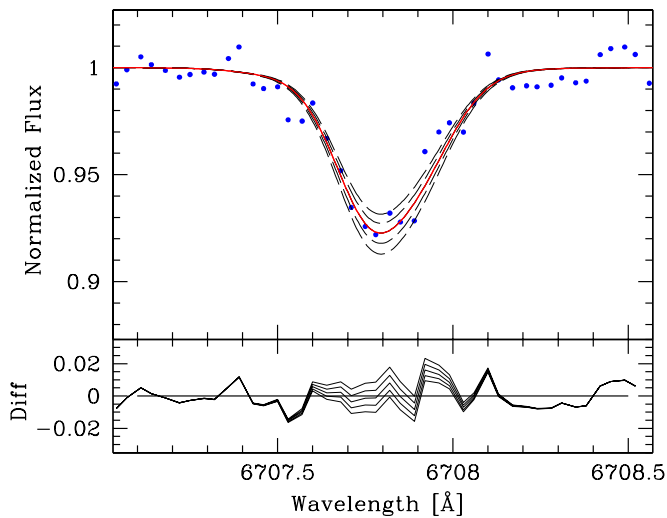}}
\caption{The \ion{Li}{i} line at 670.8\,nm in 
MOA-2010-BLG-285S. The dots are the 
observed spectrum and the thick solid line the best fit representing 
an abundance of $\log\epsilon({\rm Li}) = 2.21$. Thin dashed lines 
represent different Li abundances from 2.15 to 2.27\,dex in 
steps of 0.03\,dex. Lower panel shows the differences between the
observed and synthetic spectra.
                  \label{fig:lifigur}}
\end{figure}

By producing a set of synthetic spectra with different Li abundances,
varying in steps of 0.03\,dex, we performed a
$\chi^2$-minimisation to find the best fitting synthetic spectrum, and
hence the Li abundance. The best fitting value is
$\log\epsilon({\rm Li})=2.21$ (see Fig.~\ref{fig:lifigur}). 
According to \cite{lind2009}
the 1-D non-LTE correction for a star with 
$\teff$/$\log g$/[Fe/H]\,$=$\,6064/4.2/$-$1.23 is $-0.05$\,dex 
for the 670.8\,nm line, giving a NLTE corrected abundance of
$\log\epsilon({\rm Li})=2.16$ for MOA-2010-BLG-285S.
We expect that the 3D non-LTE abundance would be quite similar
\citep{asplund2003b,sbordone2010}.

The uncertainty in the Li abundance has three main sources: 
the continuum level; the line profile fitting; 
and the effective temperature. Based on
the high quality of the spectrum ($S/N\approx 170$), and also by 
visual inspection of the spectrum, the uncertainty due to the 
placement of the continuum was estimated by changing the 
continuum level by 0.005
and then redoing the fitting. The uncertainty
due to the fitting is estimated to be 0.03\,dex  
(see also Fig.~\ref{fig:lifigur}). Finally we determined
a new abundance by changing $\teff$ with 129\,K (the 1$\sigma$
uncertainty, Table~\ref{tab:parameters}).
Adding the three uncertainties in quadrature (assuming that they
are uncorrelated) gives a total error of 0.10\,dex.

\begin{table}[hb]
\centering
\caption{
Stellar parameters and abundances for MOA-2010-BLG-285S\label{tab:parameters}
}
\setlength{\tabcolsep}{2.5mm}
\begin{tabular}{rl|rl}
\hline\hline
\noalign{\smallskip}
  $\rm[Fe/H]=$               & $-1.23\pm 0.09$      &   $\rm[O/Fe]= $    & $+0.52\pm 0.27$      \\
  $T_{\rm eff}=$             &  $6064\pm129$\,K     &   $\rm[Mg/Fe]=$    & $+0.42\pm 0.07$      \\
  $\log g=$                  &  $4.20\pm0.23$       &   $\rm[Si/Fe]=$    & $+0.30\pm 0.07$      \\
  $\xi_{\rm t}=$             &  $1.85\pm0.38\,\kms$ &   $\rm[Ca/Fe]=$    & $+0.35\pm 0.06$      \\
  Age=                       &  $12.5\pm3$\,Gyr     &   $\rm[Ti/Fe]=$    & $+0.38\pm 0.12$      \\
  $v_{\rm r, heliocentric}=$ & $+46.0\,\kms$        &   $\rm[Ni/Fe]=$    & $-0.08\pm 0.11$      \\
  $\log\epsilon({\rm Li})=$  &  $2.16\pm0.10$       &   $\rm[Cr/Fe]=$    & $-0.01\pm 0.19$      \\
                             &                      &   $\rm[Na/Fe]=$    & $-0.05\pm 0.05$      \\
\noalign{\smallskip}
\hline
\end{tabular}
\end{table}

\begin{figure}
\resizebox{\hsize}{!}{
\includegraphics[angle=-90,bb=260 50 525 470,clip]{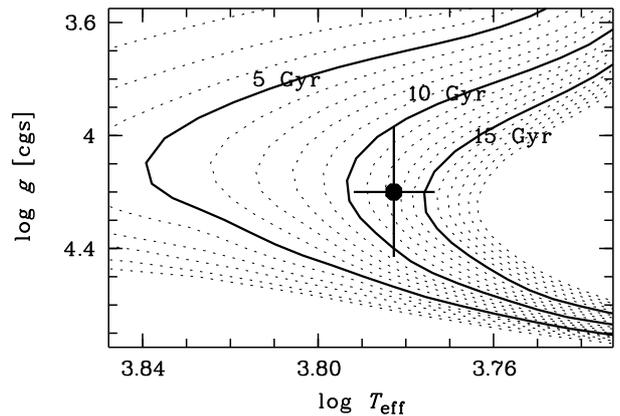}}
\caption{
         Y$^2$ isochrones \citep{demarque2004} with $\rm [Fe/H]=-1.20$
         and $\rm [\alpha/Fe]=+0.4$ that were used to 
         estimate the stellar age. 
                  \label{fig:getage}}
\end{figure}

In the previously published sample of 15 microlensed bulge dwarf stars in 
\cite{bensby2010}, we can detect the Li line in some of the stars. However,
most of them have effective temperatures well below 5900\,K, meaning 
that much of their initial Li has been destroyed 
(see Sect.~\ref{sec:intro}). Many of them also have spectra that
are of less good quality than that obtained for MOA-2010-BLG-285S, 
making Li detections even more difficult. Li abundances for some of 
the stars in \cite{bensby2010} will be included in an 
upcoming publication where we will present an extended sample 
of microlensed bulge dwarfs from the 2010 campaign.

\section{Results and discussion}

Table~\ref{tab:parameters} lists the stellar parameters
and abundance ratios for MOA-2010-BLG-285S. It is 
a warm and metal-poor turn-off star with 
an age of $12.5\pm3$\,Gyr as inferred from Y$^2$ isochrones
\citep{demarque2004} (see Fig.~\ref{fig:getage}).
At $\rm [Fe/H]=-1.23$ it is also the currently 
most metal-poor dwarf star known in the Galactic bulge, significantly
more metal-poor than the previous record holder OGLE-2009-BLG-076S
at $\rm [Fe/H]=-0.86$ \citep{bensby2009letter}. Its
$\alpha$-element abundances show enhancements of 0.3 to 0.5\,dex
relative to iron, and the abundances of the iron-peak elements Cr and Ni 
are (within the error-bars) close to solar 
(i.e., $\rm[Cr, Ni/Fe]\approx0$). This abundance pattern is 
consistent with what is seen in thick disk dwarf stars at the same 
metallicity \cite[e.g.,][]{bensby2005,bensby2007letter2,reddy2008}.

\begin{figure}
\resizebox{\hsize}{!}{
\includegraphics[bb=18 165 592 480,clip]{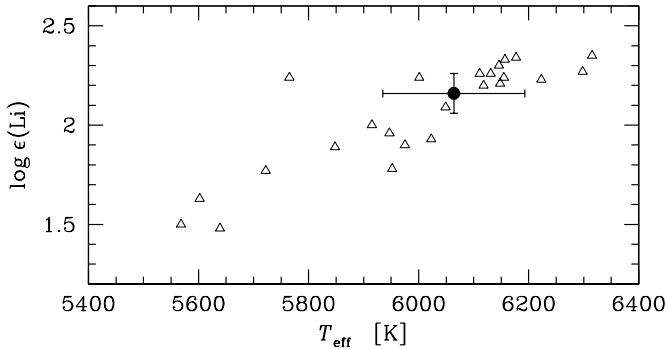}}
\caption{ 
Comparison of the Li abundance for MOA-2010-BLG-285S (filled circle)
with the stars from \cite{melendez2010li} that have $\rm -1.5<[Fe/H]<-1.0$
(open triangles).
All abundances have been corrected for non-LTE effects as given by
\cite{lind2009}.
                 \label{fig:melendez}}
\end{figure}

\begin{figure}
\resizebox{\hsize}{!}{
\includegraphics[bb=18 165 592 580,clip]{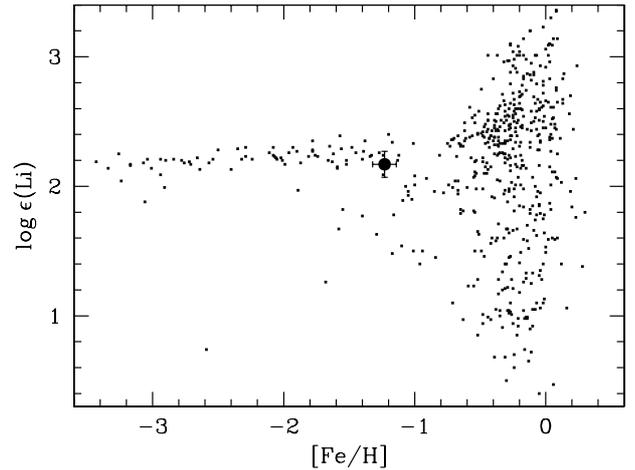}}
\caption{Li abundance versus [Fe/H]. The filled circle indicates 
MOA-2010-BLG-285S. 
Comparison data (small dots) come from
\cite{melendez2010li} and \cite{lambert2004}.
                 \label{fig:life}}
\end{figure}

MOA-2010-BLG-285S is the first metal-poor dwarf star for which Li has 
been clearly detected in the Galactic bulge. The Li abundance we 
find for MOA-2010-BLG-285S,
$\rm\log\epsilon(Li)=2.16$, is fully consistent with what is seen in 
other metal-poor dwarf stars in the Galactic disk and halo at this 
effective temperature and metallicity
(see Fig.~\ref{fig:melendez}). In Fig.~\ref{fig:life} we see that the 
star lies on the  metal-rich end of the Li Spite plateau 
\citep{spite1982}. Combined with its old age, MOA-2010-BLG-285S is an 
excellent confirmation that the bulge did not undergo a large amount 
of Li production or astration in its earliest phases.

Furthermore, the Li abundance in MOA-2010-BLG-285S,
when coupled with observations in a different galaxy ($\omega$ Cen, \citealt{monaco2010})
and different components of the Galaxy (halo, \citep[e.g.,][]{melendez2006}, 
thick disk, \citep{molaro1997}, and bulge (this work)),
suggests that the Spite plateau is universal. Also, the measured value for 
MOA-2010-BLG-285S is in agreement with models 2, 3, and 4 of
\cite{matteucci1999} that predict negligable production of Li 
by the $\nu$-process in supernovae, carbon stars and massive AGB stars
until $\rm [Fe/H]>-1$ in the bulge.

\section{Summary}

In this letter we report the discovery and analysis of the most
metal-poor dwarf star in the Bulge exposed to detailed abundance 
analysis. Had it not been for gravitational microlensing, during which 
its apparent magnitude was amplified by a factor of about 550, the star 
would be much too faint to be accessible for high-resolution 
spectroscopy. According to models of standard 
stellar evolution, the star should have retained most of its initial 
Li abundance, which enables us to compare with the predictions from 
Big Bang nucleosynthesis (BBNS). 
The only other detections of Li in the Galactic bulge are
from observations of RGB and AGB stars
\citep[e.g.,][]{gonzalez2009} in which the 
atmospheric Li abundance has been altered. Dwarf and subgiant
stars with effective temperatures greater than about 5900\,K 
are the only reliable tracers of Li (see discussion above).

Our main results are: 

\begin{enumerate}
\item 
At $\rm [Fe/H]=-1.23$ MOA-2010-BLG-285S is the currently 
most metal-poor dwarf star in the Galactic bulge. 
It has an old age of $12.5\pm3$\,Gyr
and enhanced [$\alpha$/Fe] ratios, consistent with those
seen in the Galactic thick disk.
\item 
For the first time, Li has been clearly detected in 
a metal-poor dwarf star in the Galactic bulge.  We 
find an NLTE corrected Li abundance of 
$\rm\log\epsilon(Li)=2.16$, which is in excellent agreement 
with Galactic halo and thick disk dwarf stars at the same 
effective temperature and metallicity, showing that the 
chemical similarities between the Bulge and other old populations 
extend also to Li. Its placement on the metal-rich end of the 
Spite Li plateau indicates that the Bulge did not undergo significant
enrichment of Li in its earliest phases. 
\item 
The similar Li abundances found in the bulge, halo, thick disk, and 
in another galaxy ($\omega$ Cen) suggest that the lithium
Spite plateau is universal.
\end{enumerate}

\begin{acknowledgement}

 We thank the referee Prof. Piercarlo Bonifacio for valuable comments.
 S.F. is a Royal Swedish Academy of Sciences Research Fellow supported 
 by a grant from the Knut and Alice Wallenberg Foundation. 
 Work by S.D. was performed under 
 contract with the California Institute
 of Technology (Caltech) funded by NASA through the Sagan Fellowship
 Program.
 Work by A.G. was supported by NSF grant AST-0757888.

\end{acknowledgement}

\bibliographystyle{aa}
\bibliography{referenser}

\end{document}